\documentstyle[psfig]{l-aa}     % LaTeX A&A  Standard Fonts

\def\ros{{ROSAT }}
\def\com{{COMPTEL }}
\def\etal{{\it et\,al. }}

\def\it{\sl}
\def\degs{\ifmmode ^{\circ}\else$^{\circ}$\fi}
\def\amin{\ifmmode ^{\prime}\else$^{\prime}$\fi}
\def\asec{\ifmmode ^{\prime\prime}\else$^{\prime\prime}$\fi}
\def\fss{\hbox{$.\!\!^{\rm s}$}}        % Fractions of seconds
\def\fdg{\hbox{$.\!\!^\circ$}}          % Fractions of degrees
\def\h{$^{\rm h}$}
\def\m{$^{\rm m}$}

\begin{document}
 
   \thesaurus{06         % A&A Section 6: Form. struct
              (
              02.01.2;  % Accretion, accretion disks
              13.07.1;  % Gamma rays: bursts
              )}

    \title{Rapid follow-up ROSAT observation of GRB 940301} 

   \author{J. Greiner\inst{1}, N. Bade\inst{2},
          K. Hurley\inst{3}, R.M. Kippen\inst{4}, J. Laros\inst{5}}

   \offprints{J.\,Greiner,\,jcg@mpe-garching.mpg.de}
 
  \institute{$^1$Max-Planck-Institut f\"ur extraterrestrische Physik,
             85740 Garching, Germany \\
         $^2$ Sternwarte Hamburg, 21029 Hamburg, Gojenbergsweg 112, Germany \\
         $^3$Space Science Laboratory, University of California,
               Berkeley CA 94720, USA\\
         $^4$Space Science Center, University of New Hampshire,
               Durham, NH 03824, USA\\
         $^5$University of Arizona, Department of Planetary Sciences, 
              Tucson, AZ 85721, USA\\}

   \date{Received January 3, 1996; accepted June 14, 1996}
 
   \maketitle
 
   \begin{abstract}
The strong $\gamma$-ray burst of March 1, 1994 was imaged by \com and found
to have an identical location within the errors as a burst which occurred
on July 4, 1993. This location coincidence had prompted speculations
on a possible single source origin for both bursts.

We have performed a \ros PSPC mosaic observation within four weeks of 
GRB 940301. 
The results of these observations and the comparison of the intensities of
the detected sources with those detected during the \ros all-sky-survey
in September 1990 are presented. We neither find a flaring X-ray source
in the April 1994 observation nor any variability of the X-ray sources
detected in the all-sky-survey.

We discuss the consequences of our negative result on both, the possibility
of the location coincidence being due to a repeating burst source as
well as due to two independent sources. In the former case the source
could either be a Soft Gamma Repeater similar to SGR 1806--20 and SGR 0525--66,
or a sofar unknown classical burst repeater. 
We conclude that a quiescent luminous X-ray source as is found for the
above mentioned Soft Gamma Repeaters is very unlikely to be present in the
case of GRB 930704 / GRB 940301.


   \end{abstract}
 
\section{Introduction}

During the first three years of operation the \com instrument on the
{\it Compton} Gamma-Ray Observatory has localized 18 $\gamma$-ray bursts
(GRBs). The location of two of these bursts, GRB 930704 and GRB 940301,
is consistent with the same location within the statistical uncertainties
(Kippen \etal 1995). With 
statistical 1$\sigma$ errors in the \com location of these events of 1\fdg5 
and 0\fdg5, respectively, the occurrence of two bursts with 1\fdg7 separation 
results in a 3\% probability for a random coincidence (Kippen \etal 1995).

Both bursts have also been observed by other instruments allowing an 
interplanetary network (IPN) location. While the IPN triangulation arc 
can neither prove nor dismiss the possibility of a true burst recurrence,
the combined \com and IPN location evaluation reduces the probability
of random spatial coincidence to 1.5\% (Kippen \etal 1995).

Besides a random coincidence there are two alternatives for the origin of 
these two bursts (Kippen \etal 1995): 
(1) These events may be time-delayed images of the same 
event which has been gravitationally lensed by an intervening massive object 
(Paczy\'{n}ski 1986). Using high time resolution OSSE data
and spectral information from the BATSE and \com instruments it was
convincingly shown that the gravitational lensing hypothesis 
can be ruled out due to the dissimilarities between the two GRBs
(Hanlon \etal 1995).
(2) These two events may have been produced by a single GRB source with a
separation of eight months. If the burst source resides at cosmological 
distances then many GRB scenarios have to be excluded because they invoke
singular catastrophic events such as mergers of compact objects
(Paczy\'{n}ski 1991) or failed supernovae (Woosley 1993).

   \begin{table}[ht]
      \caption{Sequence of events}
     \vspace{-0.15cm}
     \begin{tabular}{lc}
      \hline
      \noalign{\smallskip}
         ~~~~Event & Date \\
      \noalign{\smallskip}
      \hline
      \noalign{\smallskip}
       \ros all-sky survey$^{(1)}$ & Sep. 13--29, 1990 \\
       GRB 930704                  & July 7, 1993 \\
       GRB 940301                  & March 1, 1994 \\
       \ros mosaic pointings       & April 1, 1994 \\
      \hline
      \end{tabular}
      \label{events}

      \noindent{\small 
       $^{1)}$ The all-sky survey lasted 6 months; the date gives the
            time span when the location of the two GRBs was observed.}
\end{table}

Here we report on mosaic observations with the \ros position-sensitive 
proportional counter (PSPC) of the location of this pair of bursts.
The sequence of events is given in Tab. \ref{events}.
All the \ros data analysis described in the following has been performed
using the dedicated EXSAS package (Zimmermann \etal 1994).

   \begin{table*}[th]
      \caption{\ros X-ray sources from the mosaic pointings. The X-ray error 
      depends on the off-axis angle of the source in the detector.}
     \begin{tabular}{rcccrrccc}
      \hline
      \noalign{\smallskip}
     No. & Name & Position & Count- & Hardness~~ & ML~ & Identification 
                             & m$_{\rm B}$ & Optical Position \\
       &  & error         & rate & Ratio~~~~          & & $^{2)}$   
                             &  ~(mag)~ & (2000.0) \\
         &      & (\asec)        & (cts/s) & HR1$^{1)}$~~~ & & & & \\
      \noalign{\smallskip}
      \hline
      \noalign{\smallskip}
  P1 & RX J0656.1+6451 &27 & 0.029 &  0.25$\pm$0.14~ & 55.7 & EBL-WK & 19.5 & 
         06\h 56\m 07\fss2 +64\degs 51\amin 27\asec \\
  P2 & RX J0659.2+6442 &23 & 0.004 &  0.35$\pm$0.33~ & 18.4 & 2 candidates & &   \\
  P3 & RX J0655.9+6439 &24 & 0.018 &  0.39$\pm$0.20~ & 27.9 & 2 candidates & &   \\
  P4 & RX J0707.0+6435 &42 & 0.632 &  0.27$\pm$0.04~ & 1088.5 & Zw VII 118 (Sy1) & 14.6 & 
         07\h 07\m 13\fss0 +64\degs 35\amin 59\asec \\
  P5 & RX J0658.6+6435 &21 & 0.006 & --0.34$\pm$0.26~ & 24.8 & HD 50630 (G5) & 8.9 & 
         06\h 58\m 39\fss2 +64\degs 35\amin 34\asec  \\
  P6 & RX J0657.6+6433 &21 & 0.004 &  0.50$\pm$0.36~ & 13.7 & 2 candidates & &   \\
  P7 & RX J0659.0+6432 &20 & 0.008 &  0.06$\pm$0.24~ & 42.1 & blue object$^{3)}$ & 18.8 & 
         06\h 59\m 01\fss0 +64\degs 32\amin 35\asec  \\
  P8 & RX J0655.5+6431 &25 & 0.004 &  1.00$\pm$0.00~ & 13.0 & LTS & 11.5 & 
         06\h 55\m 31\fss9  +64\degs 31\amin 44\asec \\
  P9 & RX J0700.6+6431 &22 & 0.003 &  0.73$\pm$0.41~ & 10.8 & Red-WK$^{3)}$ & 18.7 & 
         07\h 00\m 36\fss5  +64\degs 32\amin 09\asec  \\
 P10 & RX J0657.6+6425 &21 & 0.004 & --0.22$\pm$0.35~ & 16.0 & 3 candidates & &   \\
 P11 & RX J0652.1+6424 &35 & 0.020 &  0.15$\pm$0.39~ & 9.8 & K star & 14.5 & 
         06\h 52\m 07\fss9 +64\degs 24\amin 37\asec  \\
 P12 & RX J0658.7+6423 &20 & 0.008 &  0.73$\pm$0.20~ & 51.4 & FG star & 11.5 &  
         06\h 58\m 43\fss2 +64\degs 23\amin 39\asec \\
 P13 & RX J0659.4+6422 &20 & 0.002 &  0.23$\pm$0.47~ & 14.2 & Red-WK & 17.9--19.6 & 
         06\h 59\m 23\fss7  +64\degs 22\amin 11\asec  \\
 P14 & RX J0657.9+6419 &21 & 0.004 &  0.49$\pm$0.29~ & 38.4 & blue object$^{2)}$ & 19.8 & 
         06\h 57\m 57\fss5 +64\degs 19\amin 55\asec  \\
 P15 & RX J0701.0+6417 &22 & 0.005 &  0.59$\pm$0.30~ & 22.0 & Red-WK$^{3)}$ & 18.4  & 
         07\h 01\m 06\fss6  +64\degs 18\amin 21\asec  \\
 P16 & RX J0653.7+6413 &30 & 0.009 &  1.00$\pm$0.00~ & 10.0 & 5 candidates & &   \\
 P17 & RX J0701.7+6409 &24 & 0.006 &  1.00$\pm$0.00~ & 10.1 & 4 candidates & &   \\
 P18 & RX J0658.4+6408 &22 & 0.003 &  0.70$\pm$0.36~ & 12.9 & ?$^{4)}$ & 21.3 & 
         06\h 58\m 29\fss2  +64\degs 08\amin 49\asec  \\
 P19 & RX J0700.7+6400 &26 & 0.007 &  0.78$\pm$0.46~ & 11.2 & 5 candidates & &   \\
 P20 & RX J0647.9+6339 &20 & 0.006 &  0.64$\pm$0.56~ & 11.1 & 2 candidates & &   \\
      \noalign{\smallskip}
      \hline
      \end{tabular}
      \label{idpoin}
    
      \noindent{\small 
       $^{1)}$ The hardness ratio HR1 is defined as the 
           normalized count difference (N$_{\rm 52-201}$ - N$_{\rm 11-41}$)/
           N$_{\rm 11-201}$, where N$_{\rm a-b}$ denotes the number of counts
           in the PSPC between channels a and b.\\
       $^{2)}$ Based on the positional coincidence, the classification of
           the optical objects according to the objective prism plate spectra, 
           the hardness ratio of the X-ray emission and
           the F$_{\rm X}$/F$_{\rm opt}$ ratio. LTS denotes late-type star,
           and Sy1 means Seyfert 1 galaxy. Spectral types of the stars
            are given in parenthesis. 
           Notations of 
          ``EBL-WK" (weak blue object with emission lines) or ``Red-WK"
          (weak red object) are descriptions of the objective prism
          spectra and indicate extragalactic objects
           (see Bade \etal 1995 for details on the object classification). \\ 
      $^{3)}$ Identification uncertain. Another optical object inside error 
           circle. \\
      $^{4)}$ The optical brightness and position is given for the only 
            object in the X-ray error circle down to 
            B=22$^{\rm m}$.
           }
   \end{table*}

\section{X-ray observations}

\subsection{ROSAT PSPC Pointings}

ROSAT pointed observations of the location of the burst pair were performed
on April 1, 1994. One of the initially planned two mosaic pointings
was unfortunately split into two short exposures with slightly different
pointing directions due to scheduling problems. As a consequence, we obtained
three pointings with 3150 sec, 1020 sec and 364 sec, respectively,
which results in an uneven sensitivity over the GRB error boxes.

%______________________________________________ Gamma_1 (lg rho, lg e)
   \begin{figure*}
      \centering{
      \hspace*{.1cm}
      \vbox{\psfig{figure=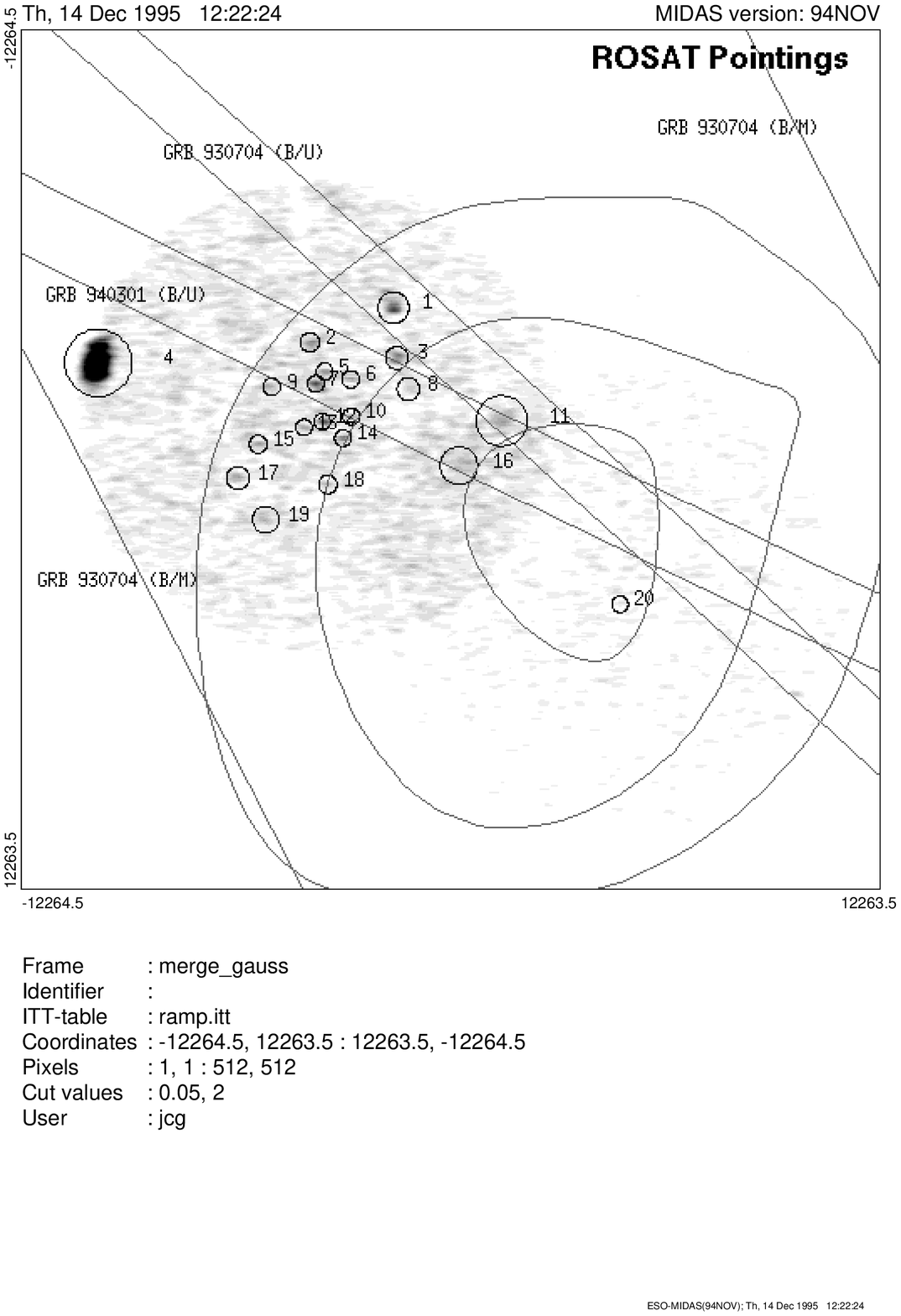,width=14cm,%
          bbllx=3.1cm,bblly=10.5cm,bburx=18.65cm,bbury=26.1cm,clip=}}\par
      }
      \caption[point]{Smoothed 3\degs\, by 3\degs\, ROSAT mosaic 
           image of the location of the GRB 930704/940301 burst pair. 
           Circles denote X-ray sources with a likelihood greater than 8
           and the numbers reference these sources according to 
           Tab. \ref{idpoin}. The circle size is proportional to the full width 
           half-maximum of the point spread function of the PSPC.
           The COMPTEL 1 to 3$\sigma$ contours of GRB 940301 
           are given as closed polygons,
           and the IPN arcs are labelled with
           the burst and B/U (BATSE/Ulysses) or B/M (BATSE/Mars Observer)
           for the satellite combinations.
           }
         \label{poin}
    \end{figure*}

Using a maximum likelihood technique we detected 19 sources in the longest
and 1 source in the 1 ksec observation with a likelihood exceeding 8
(95\% confidence, i.e. 5 out of 100 detected sources are spurius). 
The merged image of all three pointings is shown in Fig. 1, 
with the detected sources
marked by small circles and referenced with numbers detailed in 
Tab. \ref{idpoin}.
In addition to the ROSAT source name and the X-ray position error (column 3)
we give the countrate in the ROSAT PSPC 
(column 4), the hardness ratio HR1 of the detected X-ray emission (column 5, 
see table for the definition), the likelihood for the existence of a point 
source (column 6)
the optical counterpart candidates
(column 7) and the brightness (B band) and optical position for unambiguous
identifications.

The error box of GRB 930704 as determined by the BATSE/Ulysses/Mars Observer
network is not completely covered by the ROSAT pointings.
We find two sources (P1 and P11) inside or at the border of the 3$\sigma$
error box. 

The 3$\sigma$ error box of GRB 940301 determined by COMPTEL and the IPN is fully
covered by the pointings and contains nine X-ray sources including those at 
the border of the box (P2, P3, P5, P6, P7, P8, P10, P11, P16).

\subsection{ROSAT All-Sky-Survey observation}

In addition to the dedicated pointed  observations, ROSAT data are also
available which were taken about three years before GRB 930704.
The location of the burst pair was scanned during the ROSAT all-sky-survey
between September 9 and 28, 1990 with a mean
exposure time between 350--420 sec over the 6\degs\, by 6\degs\, field.
Using a maximum likelihood method we detected 29 X-ray sources in this field
with a likelihood larger than 8 corresponding to a minimum detectable
countrate of $\approx$0.01 cts/s. These sources are shown in Fig. 2
together with the COMPTEL and IPN location contours, and  details
of these sources 
%together with optical identifications of some of them
are given in Tab. \ref{idsurv}.

%___________________________________ Two column table (place early!)
   \begin{table*}
      \caption{\ros X-ray sources from the all-sky survey observations.
      The error of the X-ray positions is identical for all sources due 
      to the scanning mode (30\asec). Source S13 is identical to source
      P4 in Tab. \ref{idpoin}. All other sources are different from those in 
      Tab. \ref{idpoin}.}
     \begin{tabular}{rccrrccc}
      \hline
      \noalign{\smallskip}
     No. & Name & Countrate & Hardness~~~~ & ML & Identification & m$_{\rm B}$ & Optical Position \\
          &   & (cts/s) & Ratio  HR1$^{1)}$ &  &  $^{2)}$  &  ~(mag)~ & (2000.0) \\
      \noalign{\smallskip}
      \hline
      \noalign{\smallskip}
  S1 & RX J0644.1+6702 & 0.058 &  0.91$\pm$0.20~ & 54.0 & 7 candidates &  &      \\
  S2 & RX J0648.5+6639 & 0.041 &  0.60$\pm$0.34~ & 25.1& LTS & 10.5 & 
         06\h 48\m 35\fss9  +66\degs 39\amin 12\asec \\
  S3 & RX J0701.7+6617 & 0.026 &  0.38$\pm$0.40~ & 9.4 & 4 candidates &  &      \\
  S4 & RX J0717.4+6603 & 0.073 & --0.04$\pm$0.20~ & 54.0 & K star & 13.0 & 
         07\h 17\m 29\fss1  +66\degs 03\amin 39\asec \\
  S5 & RX J0719.2+6557 & 0.165 &  0.48$\pm$0.12~ & 126.1 & AGN or QSO &  &      \\
  S6 & RX J0636.3+6554 & 0.196 & --0.96$\pm$0.05~ & 162.9 & blue object $^{3)}$& 19.5 &
         06\h 36\m 22\fss7 +65\degs 54\amin 15\asec \\
  S7 & RX J0642.0+6552 & 0.015 &  1.00$\pm$0.00~ & 9.7 & 4 candidates &  &      \\
  S8 & RX J0720.7+6543 & 0.023 &  1.00$\pm$0.00~ & 15.0 & M star & 16.7  &
         07\h 20\m 41\fss5  +65\degs 43\amin 19\asec \\
  S9 & RX J0701.0+6541 & 0.046 &  0.29$\pm$0.27~ & 24.5 & K star & 13.5 & 
         07\h 01\m 02\fss2  +65\degs 41\amin 50\asec \\
 S10 & RX J0711.0+6533 & 0.015 &  1.00$\pm$0.00~ & 8.8 & K star & 16.5 &
         07\h 11\m 03\fss9  +65\degs 34\amin 10\asec \\
 S11 & RX J0648.4+6520 & 0.018 &  1.00$\pm$0.00~ & 12.8 & LTS & 11.7 & 
         06\h 48\m 27\fss6  +65\degs 20\amin 57\asec  \\
 S12 & RX J0658.1+6516 & 0.031 & --0.30$\pm$0.39~ & 9.5 & HD 50452 (F5) & \,~9.1 & 
         06\h 58\m 09\fss3  +65\degs 16\amin 20\asec \\
 S13 & RX J0707.2+6435 & 0.858 &  0.28$\pm$0.06~ & 998.2 & Zw VII 118 (Sy1) & 14.6 & 
         07\h 07\m 13\fss0 +64\degs 35\amin 59\asec \\
 S14 & RX J0717.8+6430 & 0.036 &  0.70$\pm$0.23~ & 23.6 & QSO & 17.7 & 
         07\h 17\m 53\fss9 +64\degs 30\amin 48\asec     \\
 S15 & RX J0638.9+6408 & 0.027 &  0.36$\pm$0.35~ & 13.3 & HD 46606 (K2) & \,~8.9 & 
         06\h 38\m 57\fss1  +64\degs 09\amin 23\asec \\
 S16 & RX J0642.7+6405 & 0.040 &  0.17$\pm$0.30~ & 18.6 & HD 47373 (K0) & \,~8.9 & 
         06\h 42\m 46\fss2  +64\degs 05\amin 46\asec \\
 S17 & RX J0632.7+6340 & 0.304 &  0.92$\pm$0.04~ & 370.6 & MCG+11-08-054 & 13.3 &  
         06\h 32\m 47\fss9 +63\degs 40\amin 25\asec \\
 S18 & RX J0705.4+6333 & 0.079 &  0.15$\pm$0.22~ & 43.9 & 2E\,0700.7+6338 (Sy1) & 15.4 &
         07\h 05\m 29\fss3  +63\degs 33\amin 32\asec \\
 S19 & RX J0704.3+6318 & 0.191 &  0.98$\pm$0.08~ & 71.2 & Galaxy $^{4)}$ & 17.2 & 
         07\h 04\m 23\fss6 +63\degs 18\amin 30\asec \\
 S20 & RX J0652.2+6246 & 0.030 & --0.68$\pm$0.38~ & 8.1 & M star$^{5)}$ & 15.3 &
         06\h 52\m 16\fss8  +62\degs 47\amin 10\asec \\
 S21 & RX J0642.3+6228 & 0.021 &  1.00$\pm$0.00~ & 10.5 & K star & 13.3 & 
         06\h 42\m 16\fss9  +62\degs 28\amin 49\asec \\
 S22 & RX J0712.4+6216 & 0.021 &  1.00$\pm$0.00~ & 17.4 & HD 54318 (B9)$^{6)}$ & 
  \,~8.0 & 07\h 12\m 28\fss0 +62\degs 15\amin 46\asec  \\
 S23 & RX J0657.9+6219 & 0.304 &  0.05$\pm$0.10~ & 192.7 & LHS 1885 (M) & 15.6 &  
         06\h 57\m 55\fss0  +62\degs 19\amin 42\asec \\
 S24 & RX J0704.0+6214 & 0.040 &  0.43$\pm$0.43~ & 15.6 & K or M star &  &      \\
 S25 & RX J0714.8+6208 & 0.046 &  0.02$\pm$0.29~ & 25.7 & HD 54943 (G0) & \,~8.5 &   
          07\h 14\m 54\fss0  +62\degs 08\amin 14\asec \\
 S26 & RX J0659.0+6212 & 0.019 &  1.00$\pm$0.00~ & 10.5 & FG star$^{5)}$ & 14.4 & 
         06\h 58\m 59\fss7  +62\degs 13\amin 17\asec \\
 S27 & RX J0655.5+6211 & 0.053 &  0.24$\pm$0.30~ & 16.6 & HD 50054 (G5) & \,~9.2 & 
         06\h 55\m 38\fss1  +62\degs 11\amin 32\asec \\
 S28 & RX J0704.1+6203 & 0.104 &  0.08$\pm$0.19~ & 64.0 & red object$^{7)}$ & 20.2 &
         07\h 04\m 09\fss9  +62\degs 03\amin 27\asec \\
 S29 & RX J0708.7+6135 & 0.057 & --0.35$\pm$0.25~ & 28.4 & M star & 16.5  & 
         07\h 08\m 45\fss0  +61\degs 35\amin 19\asec \\
      \noalign{\smallskip}
      \hline
      \end{tabular}
      \label{idsurv}

      \noindent{\small 
       $^{1)}$ and
       $^{2)}$ See corresponding notes below Tab. \ref{idpoin}. \\
       $^{3)}$ This blue object (B--V$<$--0.5) is the only one in the 
            X-ray error circle. The extreme softness in X-rays together with
            the F$_{\rm X}$/F$_{\rm opt}$ ratio suggests a magnetic cataclysmic
            variable (AM Her type) classification. \\
       $^{4)}$ The X-ray emission is extended, suggesting one of the three
              galaxies within the X-ray error box to be the counterpart.
              The most probable object (due to brightness and orientation)
              is given in the fifth and sixth column. \\
       $^{5)}$ Identification uncertain. Other object(s) in the X-ray 
              error circle.\\
       $^{6)}$ This is the only object in the X-ray error box,
            though late B stars usually are thought to be X-ray quiet
            (Bergh\"ofer and Schmitt 1994). Either the X-rays are produced
            by an optically invisible companion or the interaction with it,
            or the optical counterpart would by very faint (below B=22 mag)
            implying a ratio of L$_{\rm X}$/L$_{\rm opt}$ larger than 20.\\
       $^{7)}$ Most of the X-ray photons came in a flare, thus this red 
            object (B--V=2.3) is probably a M star.
        }
   \end{table*}

There is one X-ray source located in (or near) each error box, 
namely S8 (GRB 940301)
and S12 (GRB 930704).  Also, there are some additional 
sources within a few arcmin distance of the GRB error box (S9, 
S13, S15, S16). 

%______________________________________________ Gamma_1 (lg rho, lg e)
   \begin{figure*}
      \centering{
      \hspace*{.1cm}
      \vbox{\psfig{figure=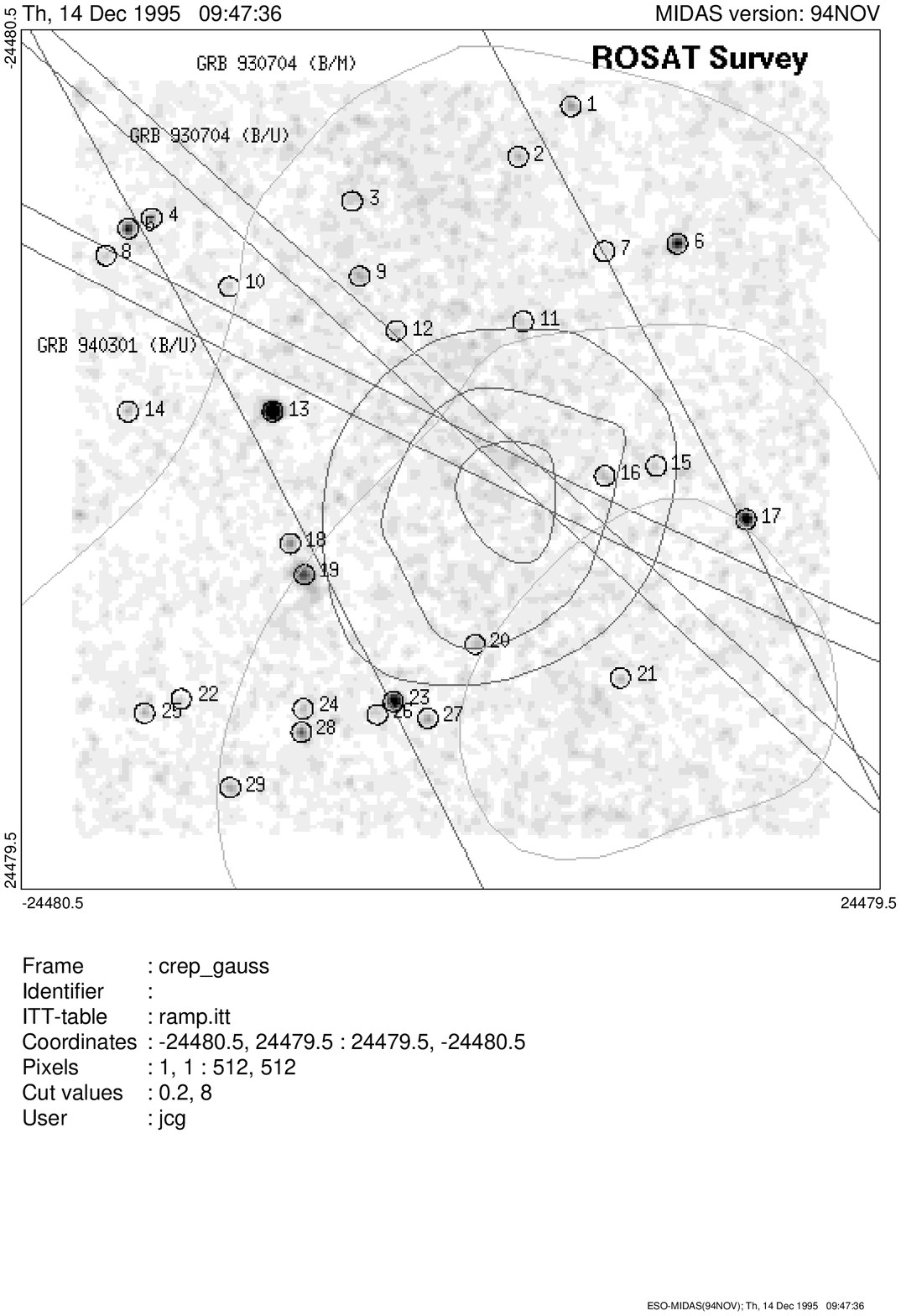,width=14cm,%
          bbllx=3.1cm,bblly=10.5cm,bburx=18.65cm,bbury=26.1cm,clip=}}\par
      }
      \caption[survey]{Smoothed 6\degs\, by 6\degs\, ROSAT all-sky-survey 
           image of the location of the GRB 930704/940301 burst pair. 
           Circles denote X-ray sources with a likelihood greater than 8
           and the numbers reference these sources according to 
           Tab. \ref{idsurv}.  The COMPTEL 1 to 3$\sigma$ contours are given as 
           bold polygons near the center of the image for GRB 940301,
           and grey polygons for the less well localized GRB 930704, 
           and the IPN arcs are labelled with
           the burst and B/U (BATSE/Ulysses) or B/M (BATSE/Mars Observer).
           }
         \label{surv}
    \end{figure*}

\subsection{X-ray variability}

The only source detected in both, the all-sky-survey scanning and the pointed
observation in April 1994, is the Seyfert 1 galaxy Zw VII 118 (= S13 = P4). 
Within the errors, this source is found to be constant between the two 
\ros observations.

All other X-ray sources detected in the \ros mosaic pointings 
(in particular all nine objects inside the GRB 940301 error box as 
well as the two sources inside the GRB 930704 error box) have 
intensities which are below the sensitivity threshold
of the all-sky survey observation. This explains their non-detection
in the all-sky survey data. As a consequence we can therefore only 
state that the pointed \ros observations performed four weeks after 
GRB 940301 did neither reveal any flaring nor any fading X-ray source as 
compared to the all-sky survey observation in 1990.

\section{Optical data}

The error regions of the X-ray sources were investigated at optical wavelengths
using two existing data sets: 
(1) Photographic plates with objective prism spectra taken with the 
Hamburg Schmidt telescope on Calar Alto.
In the Hamburg objective prism survey (Hagen \etal 1995) spectra are
taken in the 3400--5400 \AA\, range with a dispersion of 1390 \AA/mm down to
B $\approx$18.5. This survey covers the whole northern hemisphere 
for galactic latitudes ($|$bII$|$ $>$20\degs). 
(2) The digital version of the Palomar Observatory Sky Survey (POSS)
was used to check for the existence of objects fainter than B $\approx$18.5
as well as the colours of these objects.

The identification of the X-ray sources was done according to
the positional correlation of the optical object with
the X-ray position (which is accurate to typically less than 30\asec),
the spectral information from the objective prism plates and using the
knowledge of specific
X-ray to optical intensity ratios for different classes of objects.

Among of the X-ray sources detected during the pointed observation and
located near the GRB error boxes, two have rather secure identifications:
P11 is very probably a late-type star of spectral class K, and
P1 has a 19th mag blue object inside the X-ray error circle
with clearly detected emission lines which suggests a quasar identification. 
P1 is the second brightest source in the pointings and
is just at the detection threshold for the 350 sec survey exposure time
at this location. Though the 
optical (spectroscopic) identification is not complete, all of these X-ray 
sources inside or at the border of the GRB error boxes
have at least one object with reasonable optical brightness inside the X-ray 
error box to expect no unusual nature for the X-ray source such as a fading
GRB counterpart. 

Among the securely identified X-ray sources detected in the all-sky survey 
observation is object S8 which is a V=15 mag M star,  and S12 being
correlated with the F5 star HD 50452. S13 is the above mentioned Seyfert 1 
galaxy Zw VII 118 while the remaining other X-ray  sources are correlated 
with coronal emission from late-type (mostly K type) stars. 
Thus, none of these objects is thought to be a quiescent GRB counterpart.

\section{Discussion}

\subsection{GRB 930704 and GRB 940301 as two independent classical GRBs}

In the case of GRB 930704 and GRB 940301 being two separate bursts, the 
much wider error boxes of the  individual bursts (about 6\degs\, and 2\degs\, 
along the triangulation arcs of
GRB 930704 and GRB 940301, respectively) contain several X-ray sources.
A correlation of one of these X-ray sources to the respective burst remains
unresolved at this state. Though firm conclusions have to awit the
complete optical identification of all the X-ray sources found in the
error boxes, earlier identification work in small interplanetary network 
locations support the suspicion that none of the X-ray sources might be
related to either GRB.

\subsection{GRB 930704 and GRB 940301 from one repeating burst source}

In the case of the interpretation of the GRB 930704 / GRB 940301 burst pair 
being due to a single repeating
source (and thus the location being determined by the two crossing
BATSE/Ulysses triangulation arcs of the individual burst events), 
the source P11  is the only X-ray source
which is compatible with the burst location (just inside the 3$\sigma$
location). This source has also been detected in a 10 ksec ASCA observation
(Murakami \etal 1996). Using the ROSAT values of the countrate and hardness
ratios, and assuming a Raymond-Smith spectrum of about 1--2 keV appropriate for 
a late-type star we estimate an expected ASCA GIS countrate of 0.002 cts/s.
This is somewhat lower than the observed rate of 0.008 cts/s 
(Murakami \etal 1996), but due to the unknown spectral shape not inconsistent.
The ASCA data have the potential of checking the thermal nature of the
X-ray spectrum which is expected from a K star. 

Since the ROSAT source P11 is presumably a K star and thus not thought to 
be a GRB counterpart, we determine an upper limit for the quiescent X-ray 
emission of the hypothetical burst repeater at the position of the shortest
exposure time of 0.005 cts/s. 
At a galactic latitude of bII = 22--26\degs\, the interstellar absorption
in the direction of the burst pair is already reasonable small
(N$_{\rm H}$ = 4--7$\times$10$^{20}$ cm$^{-2}$, Dickey and Lockman 1990).
Using this absorption and a --2.2 powerlaw photon index results in
a corresponding upper flux limit of 1$\times$10$^{-13}$ erg/cm$^2$/s.

There are two alternatives about the possible nature of a bursting repeater:
(1) A Soft Gamma Repeater or (2) a classical burst which emits more than one 
burst. Though a Soft Gamma Repeater nature is already unlikely on grounds
of the observed $\gamma$-ray spectrum and temporal profile, we elaborate 
case (1) in more detail.
The quiescent luminosity of the X-ray source 
associated with the soft gamma repeater SGR 1806--20 was measured
with ASCA to be of the order of 10$^{35}$ erg/s in the 0.5--10 keV range
(Sonobe 1994). A similar hypothetical, quiescent 10$^{35}$ erg/s
source at the location of the burst pair would have to be at a distance
larger than 50 kpc, incompatible with the association with a galactic 
supernova remnant as is
generally believed to be the case for SGRs.  
Thus, the quiescent hypothetical repeater
at the location of the GRB pair is rather certainly different from the
repeater source in SGR 1806--20 (and similarly SGR 0525--66).
While one can argue that the two burst events were distinctly different
from soft repeater events in their temporal and spectral characteristics,
our data only constrain the quiescent source nature. Even if one allows 
this hypothetical quiescent repeater source to be different from the SGRs, 
then still the flaring to quiescent flux ratio of at least $\approx$10$^7$
(determined by extrapolating the GRB 940301 gamma-ray spectrum to the
ROSAT energy range with the flattest slope allowed by the COMPTEL data,
Hanlon \etal 1995) is a stringent constraint on its nature.

\subsection{The variability limit}

The fact that there is also no fading X-ray source, i.e. bright 
before the $\gamma$-ray burst and faint afterwards, has consequences
for the $\gamma$-ray  burst model involving slowly accreting neutron stars
in which a shock accompanied with the $\gamma$-ray  burst would 
prevent accretion onto the neutron star
for a time span of several years following the burst (Lasota 1992).
As detailed in Greiner \etal (1995) for the case of GRB 920622
this allows to constrain the pre-burst accretion rate of the neutron star to
(assuming a 10$^6$ K blackbody and the total galactic  absorbing column
in the direction of the GRB locations)

$$ \dot M \leq 1.5\times10^{-17}
\left[ {R \over (10 km)} \right]
\left[ M_{\footnotesize NS} \over M_{\odot}  \right]^{-1}
\left[ {D \over (100 pc)} \right]^2
M_{\odot}/yr.
%\eqno (2)
$$

Thus only for distances larger than $\approx$300 pc would the accretion rate
be high enough to trigger a hydrogen flash (Hameury \etal 1983).

\section{Summary}

Independent of the relation of GRB 930704 to GRB 940301, the pointed ROSAT
observation of the error box of the latter GRB exactly 4 weeks after the 
$\gamma$-ray event reveals no flaring/fading X-ray counterpart (upper limit of
0.005 PSPC cts/s).

While a gravitational lensing origin of the GRB 930704 / GRB 940301 pair
has been excluded already earlier (Hanlon \etal 1995), the detection
of a single repeater source is an a priori not excluded alternative
to a random spatial coincidence of two bursts.
However, this hypothetical repeater would cause problems at different
distance scales. For cosmological distances most of the scenarios
involving catastrophic events would be ruled out, whereas for galactic
distances a similar nature as the SGR sources can be excluded by our ROSAT
data. We mention that a SGR nature is also unlikely due to $\gamma$-ray
properties of both bursts.

In both cases, a lower limit on the flaring to quiescent flux ratio of
$\approx$10$^7$ has to be accomplished by a repeater source.

\begin{acknowledgements}
JG acknowledges fruitful discussions with M. Tavani during the early stage
of this project.
JG is supported by the Deutsche Agentur f\"ur
Raumfahrtangelegenheiten (DARA) GmbH under contract FKZ 50 OR 9201.
KH gratefully acknowledges Ulysses support from JPL Contract 958056 and
assistance from NASA Grant NAG5-1560.
The \ros project is supported by the German Bundes\-mini\-ste\-rium f\"ur 
Bildung, Wissenschaft,  Forschung und Technologie (BMBF/DARA) and the 
Max-Planck-Society.
This work is based on photo\-graphic data of the 
National Geographic Society -- Palomar
Observatory Sky Survey (NGS-POSS) obtained using the Oschin Telescope on
Palomar Mountain.  The NGS-POSS was funded by a grant from the National
Geographic Society to the California Institute of Technology.  The
plates were processed into the present compressed digital form with
their permission.  The Digitized Sky Survey was produced at the Space
Telescope Science Institute under US Government grant NAG W-2166.
This research has made use of the Simbad database, operated at CDS, 
Strasbourg, France.
\end{acknowledgements}

\end{document}